\documentclass[traditabstract,twocolumn]{aa}
\usepackage[varg]{txfonts}

\usepackage{graphicx}
\usepackage{natbib}
\usepackage{color}
\usepackage{subfigure}
\usepackage{xspace}
\usepackage{amssymb}
\usepackage{xspace}
\usepackage{url}
\usepackage{wasysym}

\def\fermi{{\it Fermi}-LAT\xspace}
\def\planck{{\it Planck}\xspace}

\def\grays{$\gamma$-rays\xspace}
\def\gray{$\gamma$-ray\xspace}

\def \hi{H{\sc i}\xspace}
\def \hii{H{\sc ii}\xspace}

\def \cmcube{\mbox{cm$^{-3}$}\xspace}

\def \msun{\mbox{M$_{\odot}$}\xspace}

\def \mug{$\mu$G\xspace}

\begin{document}

\title{\fermi observations of the Sagittarius B complex}
\author{Rui-zhi Yang\inst{1}
\and David I. Jones\inst{2}
\and Felix Aharonian\inst{1, 3, 4}}
\institute{Max-Planck-Institut f{\"u}r Kernphysik, P.O. Box 103980, 69029 Heidelberg, Germany.
\and Department of Astrophysics/IMAPP, Radboud University, Heijendaalseweg 135, 6525 AJ Nijmegen, The Netherlands.
\and Dublin Institute for Advanced Studies, 31 Fitzwilliam Place, Dublin 2, Ireland.
\and Gran Sasso Science Institute, 7 viale Francesco Crispi, 67100 L'Aquila (AQ), Italy.
}%

\date{Received:  / Accepted: } 

\abstract {}
{
We use 5 years of \fermi data towards the Galactic-centre giant molecular cloud complex, Sagittarius~B, to test questions of how well-mixed the Galactic component of cosmic rays are and what the level of the cosmic-ray sea in different parts of the Galaxy is.
}
{
We use dust-opacity maps from the \planck satellite to obtain independent methods for background subtraction and an estimate for the mass of the region.
We then present high-quality spectra of \gray emission from 0.3 to 30~GeV and obtain an estimate of the cosmic-ray spectrum from the region.
}
{
We obtain an estimate of the mass of the region of $1.5\pm0.2\times10^7$~\msun using the \planck data, which agrees well with molecular-line-derived estimates for the same region.
We find the the \gray flux from this region is fitted well with a cosmic-ray spectrum that is the same as is observed locally, with evidence of a small over-density at intermediate (1--10~GeV) energies.
}
{
We conclude that the \gray and cosmic-ray spectrum in the region can be well-fitted using a local cosmic-ray spectrum.
}


\maketitle

\section{Introduction}
Cosmic ray (CR) acceleration and propagation theory suggests that up to energies of $\sim10^{15}$~eV, Galactic CRs should be well mixed throughout the Galaxy.
It is these CRs that constitute the well-known {\it \emph{CR sea}}, which should describe the average density of CRs throughout the Galactic disk \citep{strong07} and can be treated as the level of the \emph{\emph{sea}} of Galactic CRs. 
However, significant questions of just how well-mixed and isotropic the CR sea is remain unanswered.
In particular, it is not clear just how strong the effect of, for example, (unknown) local CR sources \citep{Montmerle1979,yang14} or how CR mixing on intermediate (up to $\sim$a few kpc) scales affects both the normalisation and spectral behaviour of the CR sea.
As discussed below, nearby ($\lesssim1$~kpc) molecular clouds make good targets for testing the level of the CR sea (\citet{yang14} and references therein); however, variations either on intermediate size scales (e.g. a few kpc) or elsewhere in the Galaxy cannot be ruled out.

\begin{figure*}
\centering
\includegraphics[width=0.5\linewidth]{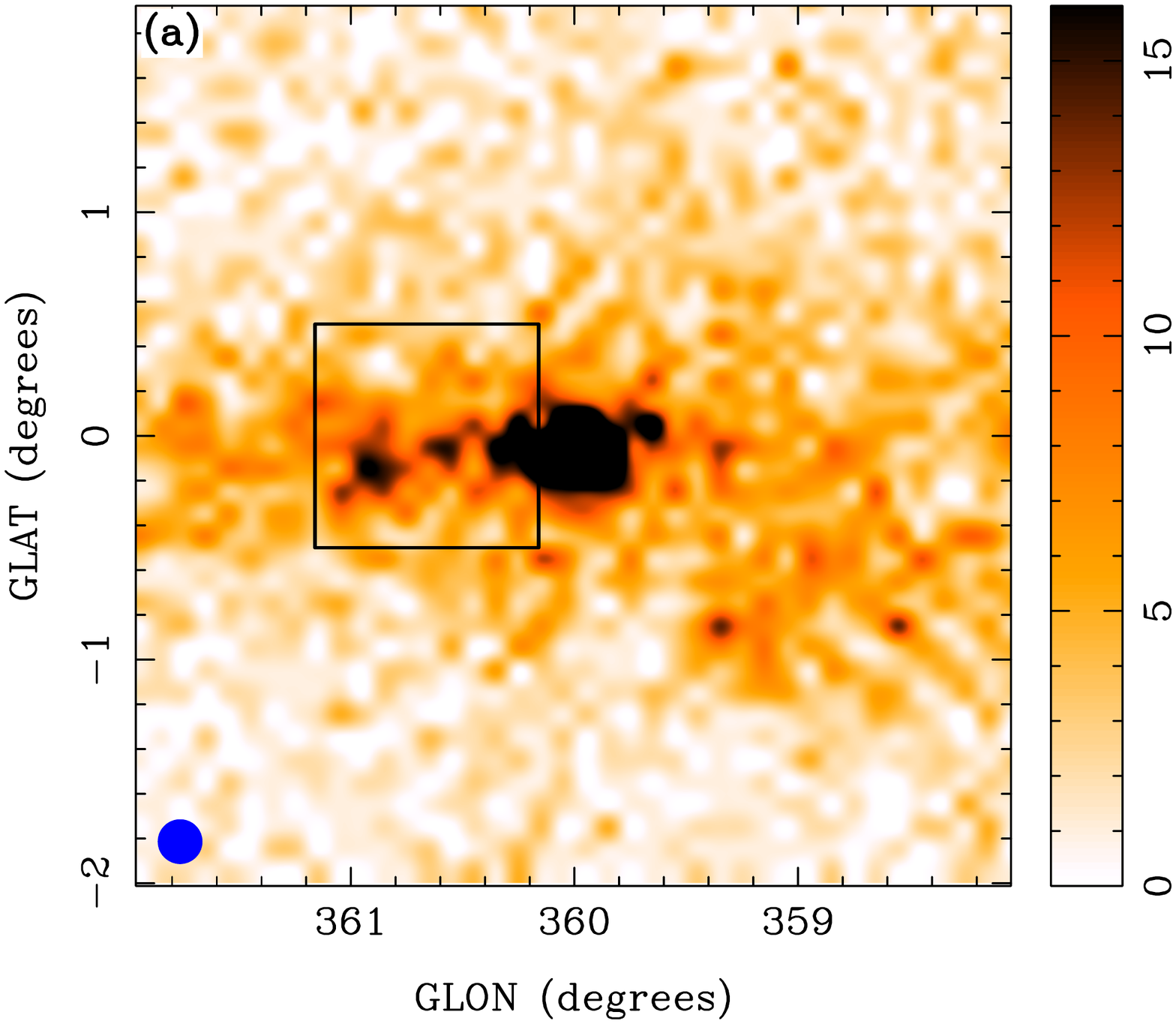}\includegraphics[width=0.5\linewidth]{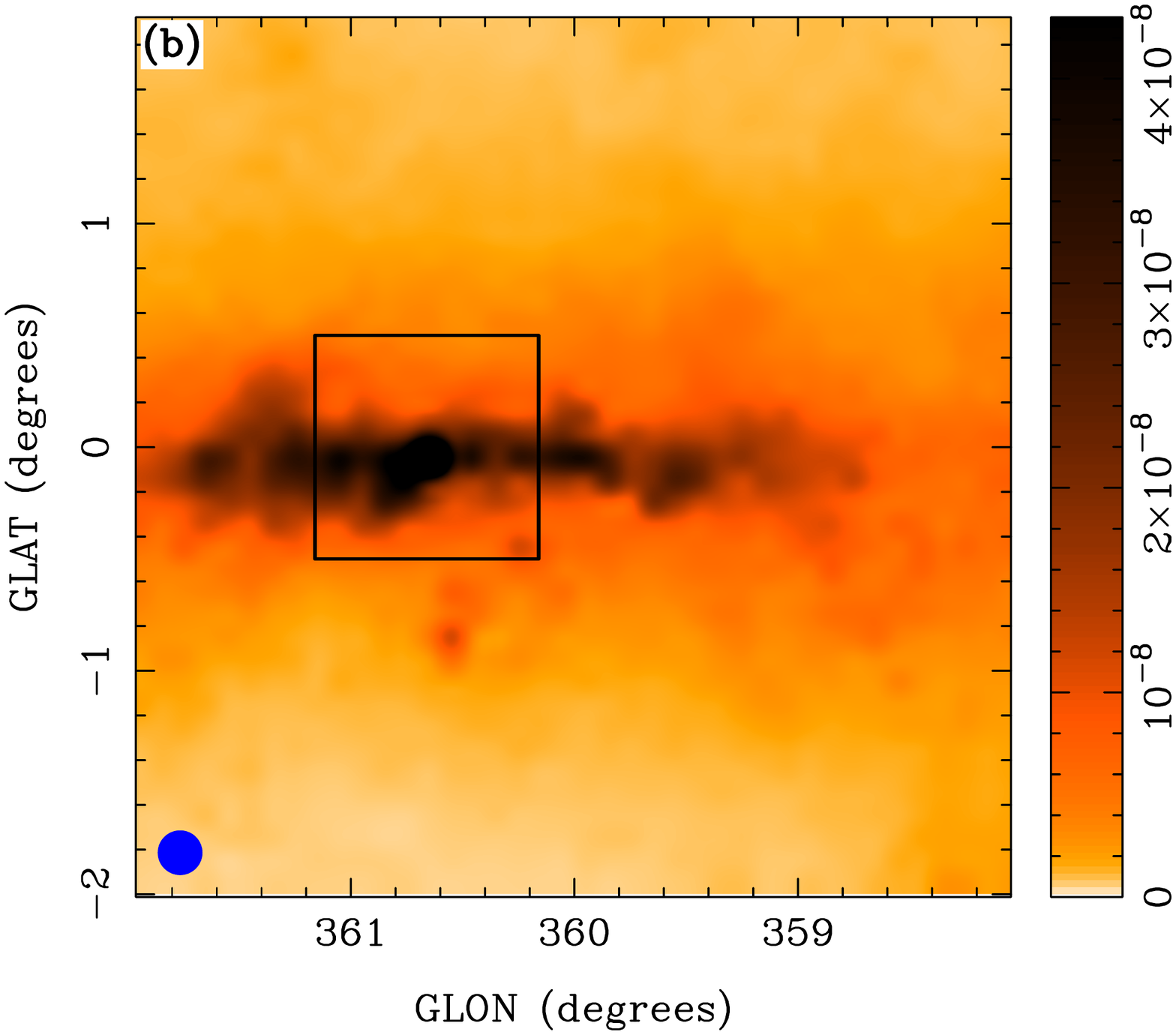}
\caption{Views of the GC region centred on the Sgr~B region (represented by the black box) and smoothed to a common resolution of $0.2^\circ$ with the beam shown in the lower left-hand corner. 
(a) \fermi count map obtained by admitting only \grays above 6~GeV, using a linear transfer function and the intensity runs from 0 to 15 counts/bin.
(b) Planck dust opacity map obtained from the \planck collaboration \citep{planck}.
This image has had a logarithmic transfer function applied to it to enhance the diffuse emission, and its scale runs from 0 to $4.5\times10^{-8}$.}
\label{fig:maps}
\end{figure*}

One of the most attractive methods of probing the structure of the CR sea is through analysing passive CR targets, such as giant molecular clouds (GMCs; e.g. \citealt{, FA91,FA2001,casanova10,pedaletti}).
However, this method requires accurate spectroscopic measurements to obtain reliable estimates to the distance $d$ and mass, $M_5$ (here $M_5$ is the mass of the GMC in units of $10^5$~\msun) of the pertinent GMC, and the ratio of these parameters, $M_5/d^2$, necessarily limits its accuracy.
Additionally, the competition of different \gray radiation mechanisms, such as pion-decay, non-thermal bremsstrahlung, and inverse-Compton (IC) losses, should also be considered in any account of the accuracy of this method. 
Nevertheless, previous studies \citep{nero12,fermiorion,fermimc,yang14} using this method have been performed on the nearby GMCs in the Gould Belt, and they confirm that the level of the CR \emph{\emph{sea}} within 1~kpc tis nearly identical to the local measurement. 
However, the $1/d^2$ relation of the \gray luminosity has limited our ability to obtain valuable information about the CR-sea in regions farther afield than these clouds.

To compensate for the decreased \gray luminosity, clouds of increased mass will also serve as ideal candidates to probe the level of the CR sea in other regions of the Galaxy.
The Sagittarius B (Sgr B) complex, which is one of the most active and massive molecular cloud complexes in the Galaxy (viz. $\sim10^7$~\msun; \citealt{tsuboi}), represents an ideal target for improving our knowledge of the CR sea.
That this cloud is located about $\sim100$~pc from the Galactic centre (GC) adds importance to observing this region.
Indeed, although the distance to the GC is large, the ratio $M_5/d^2$ is still about $10^5$, although a little less than for the better candidates in the Gould Belt, and the smaller angular size of this source results in even higher signal-to-noise ratio. 

On the other hand, because of the GC's high gas and dust content ($\sim10$\% of the total gas content of the Galaxy in $\sim1$\% of the volume over the inner $\sim400$~pc; \citealt{Crocker2011}), the high areal supernova rate (SNR) (viz. $\sim0.4$/century; \citealt{Crocker2011}), and the high total magnetic field strength of $\sim130$~\mug \citep{Crocker2010}, it is too complex and difficult a region for drawing definite conclusions.
Additionally, it is believed to be a potential source of Galactic CRs \citep{ginzburg76,ptuskin81}, so one should exercise a modicum of caution when analysing observational results from this region. 

The determination of the CR density near GC would be important for understanding the particle acceleration and origins of Galactic CRs.
This motivated us to analyse the \fermi data for the GC region.
With the \gray data from $300$ MeV to $30$ GeV and a new parametrisation for \gray production due to neutral pion decay in proton-proton collision \citep{kafexhiu14}, we derive the CR spectrum. 
The derived CR spectrum is consistent in shape with the direct measurements of local CRs by the PAMELA \citep{pamela} at the top of the Earth's atmosphere.

We have structured the paper as follows. In Section~\ref{sec:results}, we present the results of our analysis of the \fermi observations and discuss the observational uncertainties of our procedure. In Section~\ref{sec:discuss} we discuss the \gray flux from the region, as well as our derivation of the mass from the dust opacity maps and derivation of the CR spectra and fluxes, assuming that \grays are produced in interactions of CR protons and nuclei (hereafter simply CRs) with the ambient gas and finally discuss the implications of the results.
We present our conclusions in Section~\ref{sec:conc}.

\section{Results}\label{sec:results}
\subsection{\fermi data analysis}
We selected observations for which the \fermi detector was pointed towards the Sgr~B complex (MET 239557417 -- MET 408156395), which contains data over a period of approximately five years.
For this analysis, we used the standard LAT analysis software package \emph{v9r32p5}\footnote{\url{http://fermi.gsfc.nasa.gov/ssc}}. 
Given the crowded nature of the region and to avoid systematic errors due to poor angular resolution and uncertainties in the effective detection area at low energies, we selected only events with energy exceeding 300~MeV. 
The region of interest (ROI) was selected to be a $10^ \circ \times 10^ \circ$ square centred on the position of GC. 
To reduce the effect of the Earth albedo background, we excluded from the analysis the time intervals when the Earth was in the field of view (specifically when the centre of the field of view was more than $52^ \circ$ from zenith), as well as the time intervals when parts of the ROI had been observed at zenith angles $> 100^ \circ$. 
The spectral analysis was performed based on the P7rep\_v15 version of post-launch instrument response functions (IRFs). Both the front and back converted photons were selected.

\begin{figure*}
\centering
\includegraphics[width=0.5\linewidth]{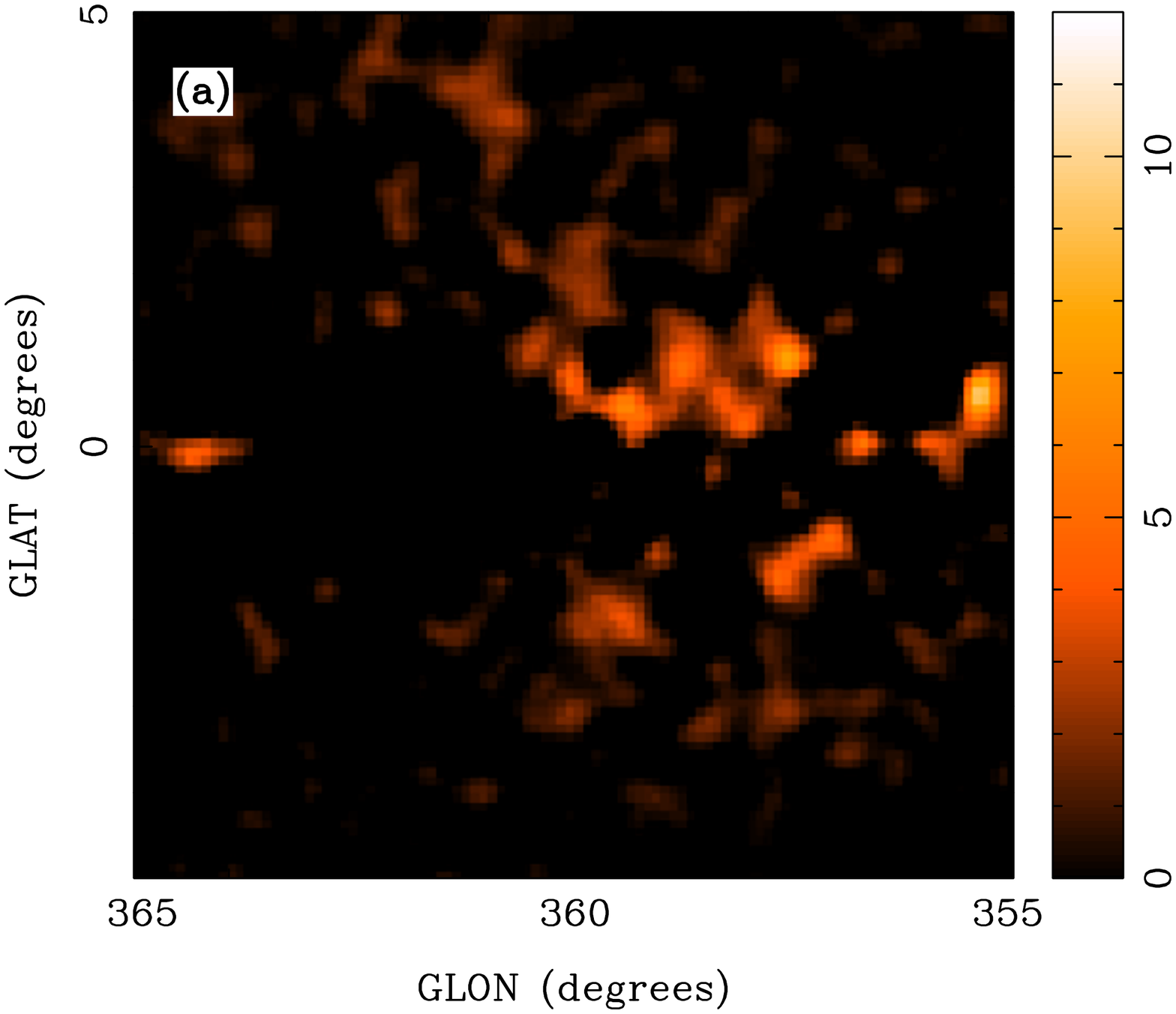}\includegraphics[width=0.5\linewidth]{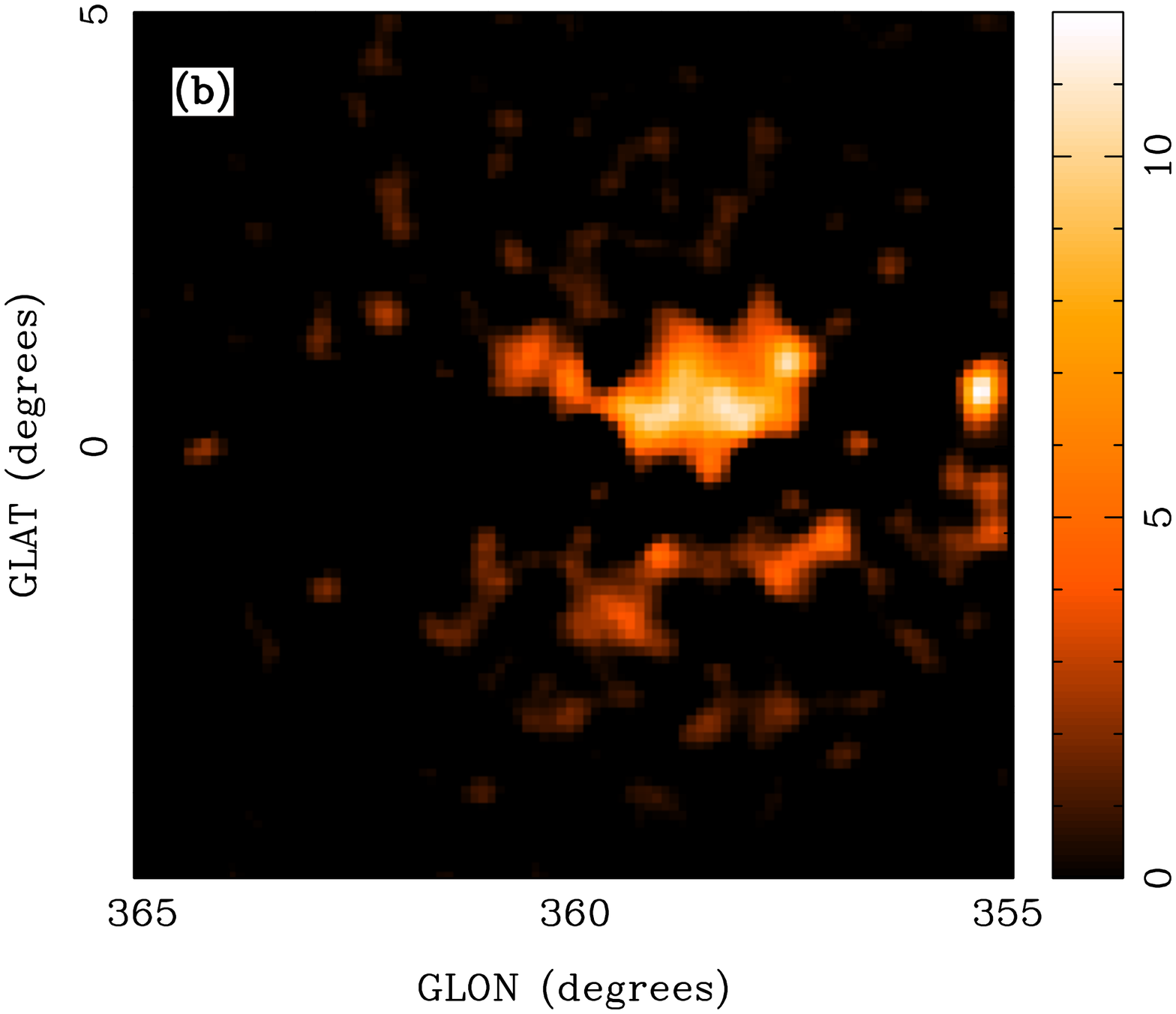}
\caption{Images of residual \grays above 1~GeV obtained using (a) the standard Fermi diffuse background (gll\_iem\_v05\_rev1.fit and iso\_source\_v05\_rev1.txt) and (b) our dust  and IC model templates, as discussed in the text.
}
\label{fig:res1}
\end{figure*}

\subsection{Background subtraction}
Since the \grays produced in molecular clouds are already included in the galactic diffuse model provided by the Fermi collaboration\footnote{gll\_iem\_v05\_rev1.fit and iso\_source\_v05\_rev1.txt, available at \url{http://fermi.gsfc.nasa.gov/ssc/data/access/lat/BackgroundModels.html}}, it cannot be used to evaluate the background. 
The Galactic diffuse emission modelled by the GALPROP software includes the contributions from the inverse Compton (IC) scattering off soft high-energy electrons, as well as the neutral pion decay and non-thermal bremsstrahlung that take place in \hi and \hii regions.
However, the contribution of non-thermal bremsstrahlung emission in passive clouds is expected to be relevant only below 100~MeV \citep[e.g.][]{gabici07}, when the electron-to-proton ratio is $e/p>$0.1. 
Considering the typical estimated ratio of $e/p\sim\rm 0.01$ as determined in observations of cosmic ray abundances at Earth \citep[e.g.][]{hillas05}, the bremsstrahlung contribution to the \gray emission can be safely neglected when modelling passive clouds. 
Thus we calculate background emission contributions only for IC using GALPROP\footnote{\url{http://galprop.stanford.edu/webrun/}} \citep{galprop}, which uses information regarding CR electrons and interstellar radiation field (ISRF). 
Isotropic templates related to the CR contamination and extragalactic \gray background are also included in our analysis.
To derive the neutral pion decay \gray emission, we used templates generated from dust opacity maps derived by the \planck collaboration \citep{planck}, where we assumed that \grays trace the spatial distribution of the molecular gas.  
It should be noted that the \planck maps have an angular resolution $<$ $0.1^\circ$ is better than, or similar to, the resolution of \fermi at all energies, so these maps may be used as templates in the \fermi analysis.
For the purposes of our analysis we chose -- as in  \citet{Crocker2007} -- to define the Sgr~B region as $0.4^{\circ}<l<0.9^{\circ},-0.3^{\circ}<b<0.2^{\circ}$ and  have used the dust opacity map in this box as our source template.
We then regarded the other regions of the dust opacity maps as background.
Figure~\ref{fig:maps}(a) shows the resulting \fermi count maps obtained from \grays with energies above 6~GeV, alongside the \planck dust opacity map Figure~\ref{fig:maps}(b), which has been smoothed using the same Gaussian kernel of $0.2^\circ$ that the \fermi data possess. 

\subsection{Point-source subtraction, likelihood analysis, and associated errors}
 We used the 3FGL catalogue \citep{3fgl} to search for point sources that fall within our region of interest and include these sources in our analysis.
The normalisation of the point sources inside the inner circle and all the diffuse templates were left as free parameters in the likelihood analysis; however, the parameters of the point sources outside the ROI were fixed to their catalogue values.  

To validate our diffuse background model, we first derived a dust template that corresponds for the whole ROI, a $15^ \circ \times 15^ \circ$ square centred on the GC. 
It should noted that the dust templates are, in fact, larger than the ROI. 
This is to take the larger PSF and possible contamination from the \grays produced outside ROI at low energies into account.  
We applied gtlike above 1~GeV both with the our dust model (dust template + IC template) and the standard Fermi diffuse background models (gll\_iem\_v05\_rev1.fit and iso\_source\_v05\_rev1.txt), as well as all the 3FGL sources, and the residual maps for both cases are shown in Figure~\ref{fig:res1}(a) and (b), respectively. 
The residual map of our model shows no clear excess compared with the standard model; however, both residuals show some excess at the region $l=[357^{\circ},359^{\circ}], b=[0^{\circ},2^{\circ}]$: this residual may related to the possible Galactic-centre GeV excess (e.g. \citealt{daylan14}). 

To test this, we introduced an additional NFW template as found in \citep{daylan14}.  
The additional template improves the fit for both diffuse models and reduces the excess in the residual maps. 
A detailed investigation of this phenomena is beyond the scope of this paper because here we are only interested in the modification of the spectrum of the Sgr B region owing to the use of additional templates. 
As a result, the difference with or without this template will be regarded as an systematic errors on our result and is discussed below.

\begin{figure*}
\centering
\includegraphics[width=0.8\linewidth]{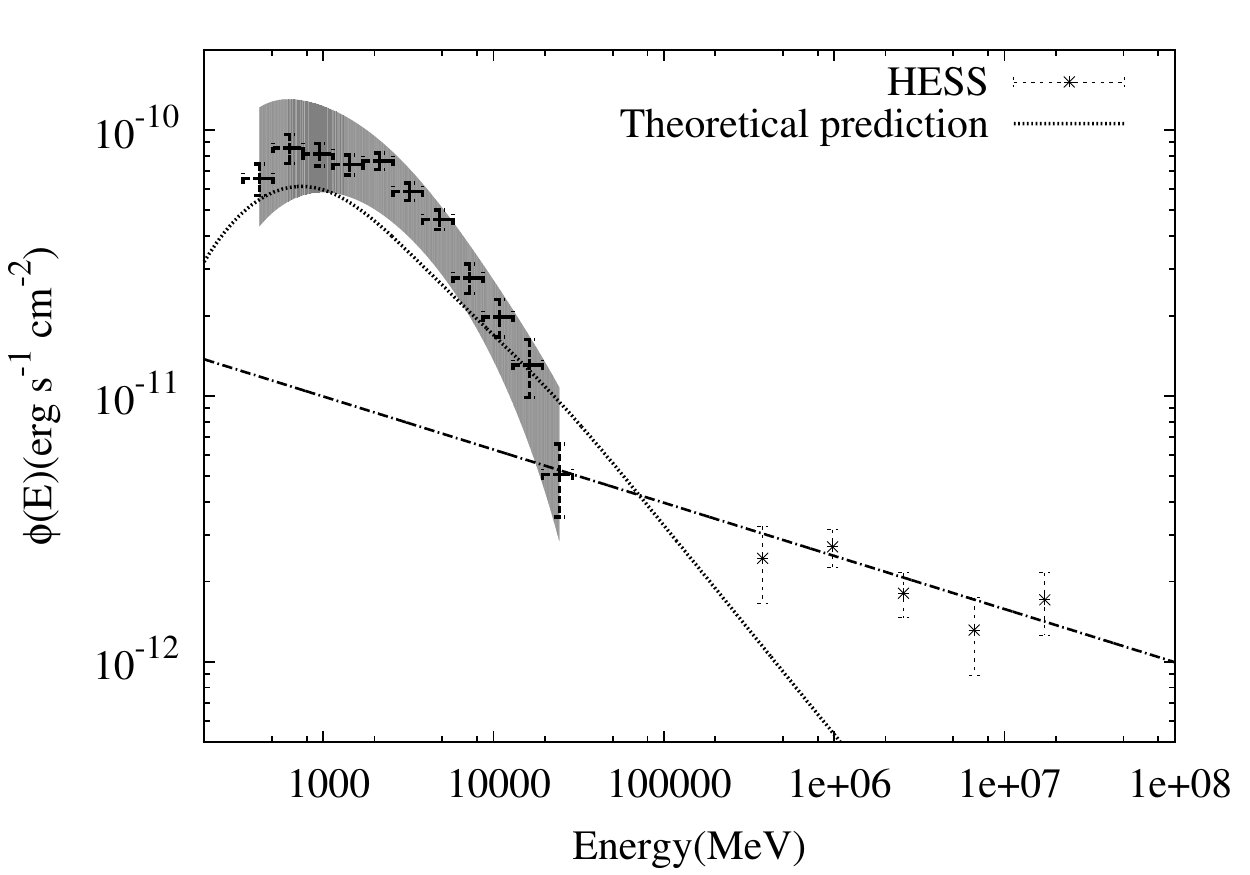}
\caption{Spectral energy distribution of \gray from the Sgr B complex. 
The grey bands represent the error of the \fermi measurement and consist of both systematic and statistic errors. 
The stars (with associated errors) represent the diffuse \gray flux discovered by H.E.S.S. \citep{hessgc}, and the dot-dashed line is the power law obtained using the photon index and normalisation obtained from the original H.E.S.S. paper reporting its discovery.
The dotted line represents the predicted \gray spectrum obtained by assuming that the CR spectrum in the Sgr~B region is the same as the local CR spectrum measured by PAMELA.
}
\label{fig:SED}
\end{figure*}

\begin{figure*}
\centering
\includegraphics[width=0.8\linewidth]{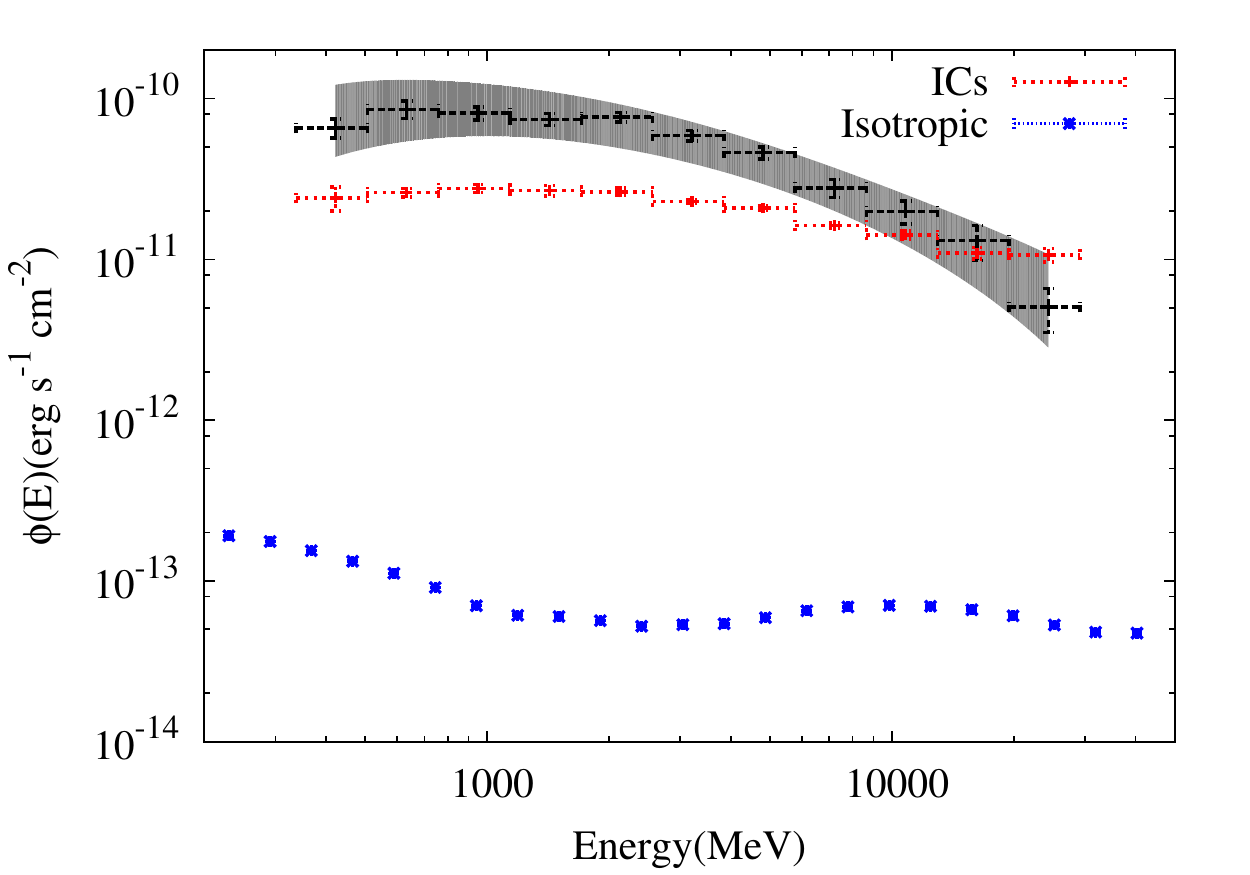}
\caption{Spectrum of different components of \gray emission from the Sgr~B region.
The (red) crosses represent the IC component, whilst the (blue) stars represent the isotropic component, both with associated errors shown.
The (black) plus signs are the total observed \gray component from the region.
}
\label{fig:bg}
\end{figure*}

To obtain the spectral energy distribution (SED) of the Sgr B complex, we divided the energy range $300~ {\rm MeV} - 30~{\rm GeV}$ into 11 logarithmically spaced bands and applied \emph{gtlike} to each of these bands. 
The results of this analysis are shown in Figure~\ref{fig:SED}. In the fitting all the 3FGL catalogue sources are kept except 3FGL  J1747.0-2828,  which is an unassociated source and which coincides with Sgr B. 
This figure shows that all the data points have a test statistic (TS) values over four, which corresponds to a significance of greater than $2\sigma$.
Apart from the systematic errors of the effective area of the \fermi, which was already taken into account before the derivation of the SED, the main source of systematic errors in the \gray spectrum comes from the uncertainty of the chosen diffuse background templates and the nearby sources, especially the central \gray source, possibly associated with Sgr~A* \citep{Chernyakova2011}. 
We used several algorithms in the likelihood fitting to investigate these (systematic) errors.
 Firstly, to investigate the systematic errors arising because of the diffuse background, we ran a global likelihood fit over the entire energy range to obtain the normalisation of the IC and isotropic templates. 
We then systematically fixed or left as a free parameter each of these templates in the likelihood fits in each energy bin, so as to observe their variation in the derived SED. 
In fact, the IC and isotropic templates only have minor influence on the spectrum for the Sgr B region. 
This is because for the Sgr B region, the \gray spectrum is dominated by emission related to the gas templates, which can be seen in Figure~\ref{fig:bg}. 
In specific regard to the central source, we attempted to fit its spectral parameters by both fixing them to the catalogue value or leaving them as a free parameter in the fit.

Another issue related to the dust templates is that the rectangular template for Sgr~B and the template for the background (dust template for the whole ROI minus Sgr~B template) all have very sharp edges. In the likelihood process of Fermi Science tool, 
however, the templates will be convolved with PSF before fitting, thus the sharp edge should not be a problem. 
Furthermore, to consider the possible leakage of the flux to or from the background templates, we also divided the background near Sgr B in smaller patches. We found with different divisions that the fits all converge successfully and the derived flux above 1~GeV is compatible. 
Only at low energies, below 1~GeV, did the different divisions cause a difference of about 10\%. 
Finally, as mentioned above, we included the NFW templates to test the possible influence of the Galactic-centre GeV excess.
The difference in the derived Sgr~B spectrum, with or without this additional template, is of the order of 20\% below 2~GeV and negligible at high energies. 

Summing all these effects, we found in the low energy bins that the systematic errors can be as large as 50\%; however, above 2~GeV, the systematic errors are compatible with the statistical errors derived in the likelihood fitting. 
The error region taking into account both systematic and statistical errors is represented in Figure~\ref{fig:SED}.
We delayed the discussion of the error that the mass estimate of the complex introduces until Section~\ref{sec:massErrors}, given that the mass does not have a bearing on the obtained \gray flux.

In addition to the resulting spectrum derived from the \fermi data from 0.1~GeV to 30~GeV, including the above-discussed errors, Figure~\ref{fig:SED} shows the data points for the diffuse TeV \grays observed by the H.E.S.S. telescope \citep{hessgc}.
Also shown in this Figure is the spectrum of the predicted \grays obtained by assuming the CR flux in this region is identical to local CR flux.
As we discuss below, it is notable that the two spectra are very similar, whilst the TeV \gray emission does not seem to match that at the lower energies explored here.

\section{Discussion}\label{sec:discuss}
\subsection{The \gray flux from the Sgr~B complex}
Figure~\ref{fig:SED} shows that the flux of MeV--GeV \grays from the Sgr~B complex is similar to the predicted \gray spectrum obtained by assuming that the CR density in the Sgr~B region is the same as PAMELA measurements. 
 We used the proton spectrum calculated by PAMELA and the $p-p$ collision cross-section parametrisation of \citet{kafexhiu14} to calculate the predicted \gray spectrum. 
For the calculation, we assumed the total mass of the gas to be $1.5 \times 10^7~\rm M_{\odot}$ (as discussed below) and used an enhancement factor of $1.79$, as suggested by \citet{kafexhiu14}, to account for the contribution from heavy nuclei.  
It should be noted that the PAMELA measurement suffered from solar modulation at low energies (mainly $E<10~\rm GeV$, see e.g. in \citealt{pamela2}). 
Thus the calculated \gray flux should be less than the prediction obtained using a demodulated CR spectrum. 
However, the uncertainties in the observed  \gray spectrum below $1~\rm GeV$ are significant, which may hinder any exact determination of the CR spectrum in this energy range. Therefore, for simplicity we do not consider the demodulation models here for simplicity.
As Figure~\ref{fig:SED} shows, at low (below 1 GeV) and high energies (above 10 GeV), the measured \gray flux is compatible with flux predicted using a local, solar-modulated CR spectrum (i.e. the PAMELA observations), whilst at intermediate energy ranges (e.g. 1--10 GeV) there is a slight excess in the measured flux, compared to what is seen locally.  
 This could be interpreted as the result of the presence of low-energy proton sources inside the Sgr B region. 
It should also be mentioned that because the PAMELA observation suffered from solar modulation in this energy range, the excess cannot reflect the true deviation between the CR density we measured and the local, demodulated CR density.
Also plotted in Figure~\ref{fig:SED} are the H.E.S.S. data points \citep{hessgc}, {\it \emph{which reveal a much harder spectrum and higher absolute flux compared with the \fermi data}}, implying a different origin for the VHE \gray emission.

\subsection{The gas content of the Sgr~B region}\label{sec:massErrors}
A major assumption in this work is that the majority of the \gray emission is produced by neutral pion decay brought about by the interaction of the CR protons with the ambient molecular material of the Sgr~B complex.
Thus a major component in the error of the CR proton flux of the Sgr~B region (see below) is its mass.
Here we derive a mass estimate based on the \planck dust opacity maps and compare it to mass estimates for the region contained in the literature.

The CS(1--0) survey of the GC using the Nobeyama 45\,m telescope by \citet{tsuboi} determined the mass of the Sgr~B complex (over a similar area to the one used here) to be $0.6 - 1.5 \times 10^7$~\msun.
 In this estimate, there is  some uncertainty about the total gas mass of the region, since the CS(1--0) traces only very dense ($10^4$~\cmcube) gas, and CO(1--0) emission, which traces lower density gas, cannot assist for both optical depth reasons and the freezing-out onto dust grains \citep{Walsh2011}. Indeed, even the robust ammonia molecule may suffer from self-absorption towards the cores of Sgr~B2 \citep{Ott2014}.%
However, other independent estimates of the total mass, as well as the major constituents of the complex, allow us to constrain the possible range of values as much as possible.
Indeed, an upper limit on the mass of the Sgr~B complex is the mass estimate of the central molecular zone (CMZ), the structure of which the Sgr~B complex is a subset. Using a range of difference tracers,
\citet{Dahmen1998} determine this mass to be $3^{+2}_{-1}\times10^7$~\msun.
The {\it Herschel} satellite also determined a mass with the 100-pc ring that it discovered of $3\times10^7$~\msun \citep{Molinari2011}.
Together, these (larger structure) estimates place a firm upper limit of $\sim3\times10^7$~\msun above which the Sgr~B complex cannot exceed.
On the other hand, the mass of the individual component, Sgr~B2, is $>6\times10^6$~\msun \citep{Jones2008}, and a crude estimate for the Sgr~B1 cloud is $\sim4.5\times10^4$~\msun \citep{Mehringer1992}.
However, \citet{Tanaka2009} found at least $5\times10^5\pm30\%$~\msun of material in a shocked region within Sgr~B1.
Thus a combined Sgr~B1 and B2 mass of $\sim10^7$~\msun -- and implying an uncertainty of a factor of about $\sim2$ -- seems more than reasonable. 

\begin{figure*}
\centering
\includegraphics[width=0.8\linewidth]{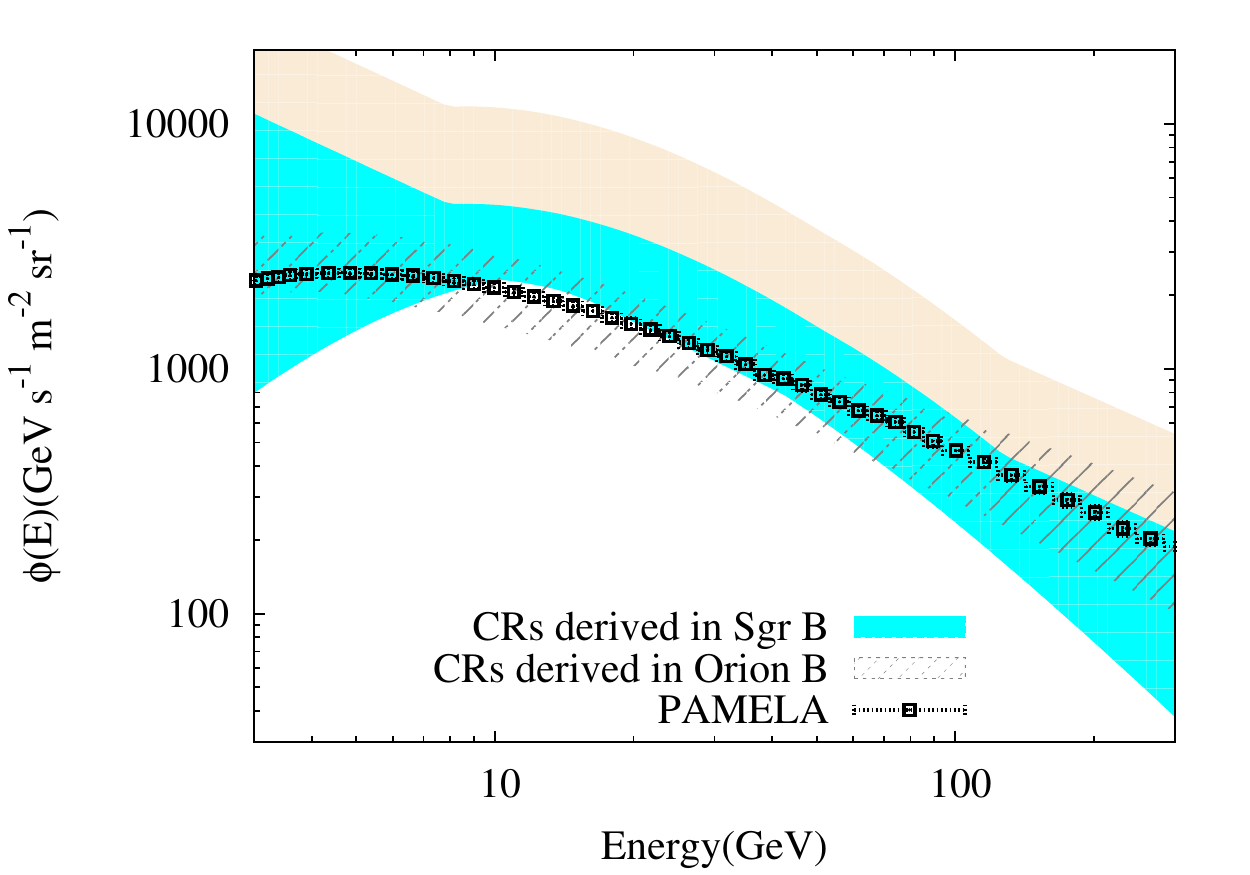}
\caption{Derived CR proton spectrum of Sgr B region as defined in the text. 
The shaded regions are derived by assuming the mass of the Sgr~B region is either $1.5 \times 10^7~\rm M_{\odot}$ (aqua) or between $0.6-1.5 \times10^7$\msun (``old lace'' white).
The crosses display the CR spectrum derived assuming a local CR spectrum, as per the PAMELA observations, whilst the grey hatched region represents the CR proton spectrum derived for the Orion~B region derived in \citet{yang14}.
}
\label{fig:SgrBprotons}
\end{figure*}

\subsection{Sgr~B Mass from the \planck dust opacity maps}
To derive the CR (proton) flux from \grays, one requires an accurate estimate of the mass, $M$, and distance, $d$\footnote{For our analysis, we fixed the distance to the Sgr B complex to $8~\rm kpc$.}, for the \gray emission region. 
As an additional cross check for the mass estimates discussed above, we have also obtained an estimate of the Sgr~B complex mass using the \planck dust opacity map shown in Figure~\ref{fig:maps} ({\it Right}).
To do this, we used a formula that relates the dust opacity and the column density using the dust as the reference emissivity according to Equation~4 of \citep{planck}:
\begin{equation}\label{eq:dust}
\tau_M(\lambda) = \left(\frac{\tau_D(\lambda)}{N_H}\right)^{dust}[N_{H{\rm I}}+2X_{CO}W_{CO}],
 \end{equation}
where $\tau_M$ is the dust opacity as a function of the wavelength, $\lambda$,  $(\tau_D/N_H)^{dust}$  the reference dust emissivity measured in low-$N_H$ regions, $W_{CO}$  the integrated brightness temperature of the CO emission, and $X_{CO}=N_{H_{2}}/W_{CO}$  the $H_2/CO$ conversion factor.
Substituting the last into Equation~\ref{eq:dust} above, one obtains
\begin{equation}
N_H = N_{H{\rm I}} + N_{H_2} =  \frac{1}{2}\tau_m(\lambda)\left[\left(\frac{\tau_D(\lambda)}{N_H}\right)^{dust}\right]^{-1}. 
\end{equation}
In our analysis we use a dust emissivity at $353~\rm GHz$ of $(\tau_D/N_H)^{dust}_{353{\rm~GHz}}=1.18\pm0.17\times10^{-26}$~cm$^2$, obtained from Table~3 of \citet{planck}. 
Thus, the total (dust-derived) mass of a cloud is
\begin{eqnarray}
 M_d&=&m_{H_2}N_{H} A_{cl} \\ \nonumber
&= &m_{H_2}\tau_D\times\left(\frac{\tau_D}{N_{H_2}}\right)^{-1} \Omega_{s}d_{kpc}^2\\ \nonumber
&\approx&2.72 \times 10^8d_{kpc}^2\int\tau_dd\Omega{\rm ~\,[\msun}],
\end{eqnarray}
where $m_{H_2}$ is the mass of the hydrogen atom, $A_{cl}$ and $\Omega_{s}$ refer to the clouds' physical and angular area respectively,  $d_{kpc}$ is the distance of the cloud in kpc, and $\int\tau_dd\Omega$  the optical depth of the dust integrated over the angular extent of the cloud in units of arcminutes$^2$. 

From this, we derive a dust mass for the entire Sgr~B complex (as defined above) of $1.5\pm0.2\times10^7$~\msun , which is consistent with the masses derived from various methods above.
Given that the mass estimate derived above is derived from the dust opacity map, which is the dust emissivity integrated over the entire line of sight, then it is no surprise that our estimate reach the upper end of the CS(1--0) mass estimate.
On the other hand, \citet{Molinari2011} used low-dust-opacity regions within the 100-pc ring region to obtain an estimate of the total dust mass from this region alone (i.e. without contamination from dust in between us and the GC, as well as behind it).
From a total mass, integrated along the line of sight, of $\sim4\times10^7$~\msun, they estimated a total 100-pc-gas-ring mass of $3\times10^7$~\msun, which is about 75\% of the total mass.
From our calculation above, 75\% of this is $1\times10^7$~\msun, which is very close to the values found by molecular line emission studies.
 In line with the fact that the Sgr~B complex contains Sgr~B2,
the most massive and active star-forming region in the Galaxy \citep{Jones2008},%
this suggests that our mass estimate is dominated purely by the material located in the GC.
However, even given the above, we still consider our estimate to be an upper limit, since this places a conservative upper limit on the CR proton flux in this region.

\subsection{CR proton spectrum from the Sgr~B C complex}\label{sec:SgrBCRs}
We have obtained an estimate of the proton spectrum in the Sgr~B complex.
This was done using the \gray observations and the parametrisation of the neutral pion decay broad band \gray spectrum introduced by \citet{kafexhiu14}. 
 The parametrisation of the cross section in \citet{kafexhiu14} possesses an accuracy of about 10\%. 
Further uncertainties may rise owing to the differences in each simulation program discussed there and may amount to 20\% in the energy range in which we are interested. 
Both uncertainties are much smaller than the uncertainties in the mass estimation.  
Detailed discussions of these uncertainties are beyond the scope of this paper, and in the calculation below we stick to the parametrisation of \citet{kafexhiu14} based on the GEANT 4 code at low energies and the PYTHIA 8.1 code at high energies.
Given the turnover in the \gray flux spectrum shown in Figure~\ref{fig:SED}, we have, in our procedure, assumed a proton spectrum to be in the form of a broken power law:
\begin{equation}
\psi(E_p)=N(E_p+E_{bk})^{-\gamma}, 
\end{equation}
where $E_p$ is the proton kinetic energy, $E_{bk}$ corresponds to the break energy, and $\gamma$ is the power-law index.  
Using a $\chi^2$ fitting routine, we find best-fit parameters for the above spectrum of $\gamma = 3.1 \pm 0.3$ and $E_{bk} = 4.5 \pm 2.0$~GeV with $\chi^2/d.o.f=2.5/8=0.3$.

Figure~\ref{fig:SgrBprotons} shows the results of our fitting procedure.
We show the results using a mass for the Sgr~B complex of $1.5\times10^7$~\msun, with the shaded region representing the $1\sigma$ error of the fitting routine.
The off-white region shows an additional part of the allowable flux-energy phase-space obtained by allowing the mass of the Sgr~B region to vary within the errors of the mass estimates (conservatively, $0.6-1.5\times10^7$~\msun; c.f. Section~\ref{sec:massErrors}).
The crosses show the spectrum resulting from fitting the PAMELA proton spectrum measurement at the top of the Earth's atmosphere, whilst the hatched region illustrates the CR proton spectrum derived by \cite{yang14} for the nearby (500~pc) Orion~B GMC.
We can see that the proton fluxes derived here vary from 1.05--4 times that of the local CR proton flux.

Considering the large uncertainties in both the measurement systematics and mass estimates, this flux is rather close to the local value.
Additionally, the proton index of $\gamma=-3.1\pm0.3$ is very steep, considering that the large-scale distribution of TeV \grays possesses a very flat spectrum of $\gamma=-2.29\pm0.27$ \citep{hessgc}.
The break energy of $E_{bk} = 4.5 \pm 2.0$~GeV is consistent with those seen in other molecular clouds and can be ascribed to propagation effects of lower-energy CR protons  presumably because of being unable to penetrate the dense regions of GMCs \citep{yang14}.

\section{Conclusions}\label{sec:conc}
We have presented new analysis of five-years worth of \fermi data.
Our analysis was motivated by fundamental, unanswered questions about the level of the CR sea in different parts of the Galaxy.
Whilst nearby clouds have been closely studied, owing to their high $M_5/d^2$ parameter, clouds that are farther away are tougher for obtaining robust \gray data.
The GMC complex of Sgr~B represents our best chance of observing the CR-sea at a distant point in the Galaxy, owing to its $M_5/d^2$ value of $10^5$, despite it being located in the GC, at a distance of $\sim8$~kpc.

We have used the 350~GHz dust opacity maps from the \planck satellite to obtain an independent mass estimate for the Sgr~B complex of $1.5\pm0.2\times10^7$~\msun.
This agrees well with independent (and varied) molecular-line estimates of the complex as a whole, as well as with molecular-line estimates from individual components within the Sgr~B complex.
However, we do note that, because the dust opacity is a measure of the total dust emissivity along the line of sight, we still consider the derived mass estimate to be an upper limit.

While the observed \gray spectrum at high energies (above 10 GeV) is compatible with the gamma-ray spectrum calculated for the local cosmic ray spectrum measured, for example, by PAMELA at low energies,  we observe some excess in the detected \gray flux at lower energies, up to
a factor of two. 
This is very different from  observations of the diffuse \gray emission of the Galactic Ridge by H.E.S.S. in the TeV domain, where the energy spectrum is much harder, and the flux is significantly  higher than the one expected from interactions of the the sea of galactic cosmic ray.  
 The slight excess of \gray flux at sub-10~GeV energies may reflect the solar modulation effects of the PAMELA measurement. 
Otherwise, it can also be explained either by the presence of low-energy particle accelerators in the Galactic entre or by effects related to the change of the character of propagation of cosmic rays (e.g. slower diffusion) of the the galactic cosmic rays in the region of the Galactic centre.
But, in either case, it is likely that the gamma-ray flux reported in this paper, at least its highest energy part, traces the sea of galactic cosmic rays.
\section*{Acknowledgements}
We are grateful to the anonymous referee, whose suggestions greatly helped to improve the manuscript.
\bibliographystyle{aa}
\bibliography{ms_lan}

\end{document}